\documentclass{acm_proc_article-sp}

\iffalse
\else
\fi

\usepackage{ucs}
\usepackage{array}
\usepackage[utf8x]{inputenc}
\usepackage{amsmath}
\usepackage[]{algorithm2e}
\usepackage{algorithmicx}
\usepackage{makecell}
\usepackage{graphicx}
\usepackage{subcaption}
\usepackage{multirow}
\usepackage{layout}
\usepackage{cite}
\usepackage{setspace}
\usepackage{listings}

\usepackage{float}

\floatstyle{ruled}
\newfloat{program}{thp}{lop}
\floatname{program}{Program}
\usepackage{color, colortbl}
\definecolor{Gray}{gray}{0.9}
\definecolor{LightCyan}{rgb}{0.88,1,1}

\definecolor{dkgreen}{rgb}{0,0.6,0}
\definecolor{gray}{rgb}{0.5,0.5,0.5}
\definecolor{mauve}{rgb}{0.58,0,0.82}

\usepackage[T1]{fontenc}
\usepackage{array}
\usepackage{makecell}
\newcolumntype{x}[1]{>{\centering\arraybackslash}p{#1}}

\usepackage{tikz}

\usepackage[T1]{fontenc}
\usepackage{mathptmx}

\begin{document}%\rmfamily%\fontsize{13}{13}
	
	\title{Energy Optimization of Source Code Guided by a Fine-Grained Energy Model}
	
	\numberofauthors{2}
	\author{
		\alignauthor Xueliang Li \qquad John P. Gallagher\\
		\email{\{xueliang, jpg\}@ruc.dk} \\
		% 2nd. author
		%\alignauthor John P. Gallagher\\
		%\email{webmaster@marysville-ohio.com}
		%
		\affaddr{Roskilde University}	\\
		\affaddr{Denmark}
	}
	\maketitle
	%{Xueliang Li \qquad John P. Gallagher\\
		%Roskilde University\\
	    %Email: \{xueliang, jpg\}@ruc.dk\\
		%\and
		%\IEEEauthorblockN{}
		%\IEEEauthorblockA{Roskilde University\\
		%Email: jpg@ruc.dk}
	%}
    %}    
	%\fi
	
	%Telephone: (800) 555--1212, Fax: (888) 555--1212}



	\begin{abstract}
		%\linespread{0.3}
Energy efficiency can have a significant influence on user experience of mobile devices such as smartphones and tablets. Software optimization plays an important role in saving energy; however,  effective energy optimization relies mainly on developers, since compiler-based energy optimizations are limited. In this paper, we propose an energy-aware programming approach,  guided by an operation-based source-level energy model, which allows the programmer to understand the energy usage of the code and then apply targeted refactoring to save energy. To the best of our knowledge, our work is the first that achieves this for a high-level language such as Java. This approach can be applied at the end of the software engineering implementation phase in order to avoid distracting developers from guaranteeing the correctness of software. In a case study, experimental evaluation shows that our approach is able to 
save from 6.4\% to 50.2\% of the overall energy consumption in various application scenarios.    

	\end{abstract}
	

	% \fi
	%\IEEEpeerreviewmaketitle

	\section{Introduction}

	Smartphones are one of the most important inventions of modern society. In February 2015, the penetration of smartphones was about 75\% in the U.S. \cite{Report:smartphonepenetration}. This figure is still growing. With the improvement of hardware processing capability and software development environments, the smartphone is no longer just a handset to make phone calls, but
also lets the user play entertaining games, watch movies, browse web pages, and so on. 
On the other hand, users are often frustrated by limited battery capacity -- applications running in parallel could easily drain a fully-charged battery within 24 hours.
	
	Software optimization by the compiler achieves very little energy-saving for mobile devices, since besides energy efficiency, the compiler for the mobile device has to consider many other
 important factors, such as limited memory usage and quick responses to user interactions. The Android platform, for instance, employs the Just-In-Time (JIT) compiler \cite{justintime}, also known as the dynamic compiler. Its optimization window is generally as small as one or two basic blocks in order to use less memory and speed up delivery of performance boost. 
However, the small window largely restricts the space of energy-saving strategies. 
More powerful code refactoring is needed, but this is beyond the scope of compilers and relies more on developers. 
	 
	%In many cases, code optimization should refer to developers.
	
	Unfortunately, current software development is performed in an energy-oblivious manner. Throughout the engineering life cycle, most developers and designers are blind to the energy usage of code written by themselves. However, developers are desperate for knowledge on energy-aware programming techniques. In the most popular software development forum \textsc{StackOverFlow} \cite{stackoverflow}, energy-related questions are marked as favorites 3.89 more often than the average questions \cite{Pinto_miningqusetion}. Furthermore, among the energy-related questions, code-design-related ones are prominent. 
	Moreover, it has been estimated that energy-saving by a factor of as much as three to five could be achieved solely by software optimization \cite{Edwards:lssmlps}. 
	To realize this, the first step is to analyze the energy attributes of source code at different levels of granularity and from different points of view.
	
	In order to expose energy attributes of code, energy modeling of code is needed to bridge the gap between high-level source code and low-level hardware, where energy is consumed. However, traditional bottom-to-top modeling techniques \cite{Tiwari:power_analysis_embedded,bran:instruction-level_model,gangqu:function-level_powermodel,Simunic:2000:source_code_optimization} face obstacles when the  software stack of the system consists of a number of abstract layers. On the Android platform, for instance, the source code is in Java and then translated to Java byte-code, further to Dalvik \cite{Android:Dalvik} byte-code, native code and machine code and finally executes on the processors and dissipates energy. Consequently, the modeling task has to describe the links between all the layers. 
	
	Instead of building a software energy model layer by layer,  
	another approach to acquiring software-level energy information is to use the hardware readings, like CPU state residency, CPU utilization, L1/L2 Cache misses and battery trace, as predictors of software energy use \cite{Dong_selfconstructivemodel,Pathak_whereisenergy,Zhang_onlinepowerestimation,Wang_batterytrace}. However, they are only capable of obtaining energy information at a coarse level of granularity such as  methods or even applications. Two pieces of work \cite{HaoShuai:2013:EstMobileApp,sourceline_energy} result in source-line energy information. The former requires low-level energy profiles. The latter employs an accurate measurement to obtain the energy dissipation of source lines. 
	
	The energy information on blocks or more coarse-grained units could identify the hot spots in the code, but it gives few clues about how to make changes to improve the code. The source line is also not an appropriate level of granularity to provide energy information. For instance,  the header of \texttt{for} loop contains three segments which are \textit{initialization}, \textit{boolean} and \textit{update} in the same source line, but usually have distinct numbers of executions. 
	%So the energy information about the source line of the header is not quite appropriate for developers.   
	
	Our latest work constructs a source-level energy model based on "energy operations", which is more fine-grained and gives more valuable information for code optimization. %Rather than coarse-grained techniques, this model can distinguish energy consumption of different operations in the same source line.   
%\footnote{To preserve anonymity in the review process, the citation is hidden}
	
By "source-level" we mean that the energy costs of running a program are all attributed to source code constructs, despite the fact that much of the energy consumed is actually accounted for by
things outside the source code such as the operating system.  Thus the model is bound to be an approximation, yet as our results show, it is precise enough to give useful information.

	Compared with coarse-grained techniques, there are some advantages of the operation-based model in guiding energy-aware programming techniques:
	
	\begin{itemize}
		\item The energy operations are basic units that constitute the energy consumption of the entire application. Thus using the energy estimate of operations,  developers can assess the effects of code changes on the energy consumption of code.  	
		%calculate the energy consumption before and after transforming the code, which directs developers on how to make changes.
		
		\item It provides more valuable information for refactoring. For example, the experiment shows that method invocation is one of the most expensive operations, suggesting that in some cases we may inline some thin methods, 
at the cost of losing the integrity of the structure of code. 
	\end{itemize}
	
	In this paper, we propose an energy-aware programming approach guided by a fine-grained energy model of source code. The summary procedure of the approach is the following:
	
	\begin{itemize}
		\item We build an operation-based source-level energy model, which is achieved by analyzing the data produced in a range of well-designed execution cases.
		\item We perform energy accounting based on the model, at operation and block level to capture the key energy characteristics of the code. 
		\item We focus efforts on the most costly blocks, where we refactor the code to remove, reduce or replace the expensive operations, while maintaining its logical consistency with the original code. 
	\end{itemize} 

	Our target platform is an Android development board with two ARM quad-core CPUs, and the source code in our study is a game engine used in games, demos and other interactive applications. We evaluate the approach in three game scenarios, and the experimental result shows that it can save energy consumption from 6.4\% up to 50.2\% depending on different scenarios.  
	
	The generality of the approach goes beyond the boundaries of the case study described here.
	Firstly, the energy-aware programming approach can be used in developing
	the large class of applications which are based on the game-engine, comprising many interactive applications with rich user interfaces.  Secondly, the approach is applicable to all kinds of applications. The choice of energy operations is dependent only on the Java source language; the techniques for designing test cases, regression analysis and code optimization can be applied to other application domains.
	
	In the rest of this paper, we begin with the identification of energy operations in Section \ref{Section_basicEnergyOp}. In Sections \ref{Section_experimentsetup} and \ref{Section:Model}, we briefly summarize (to make the paper self-contained) the setup and construction of the energy model.  Based on the model we are able to capture energy characteristics and optimize the source code in three different scenarios, \texttt{Click \& Move}, \texttt{Orbit} and \texttt{Waves}, as seen in Sections \ref{Section_clickmove}, \ref{Section_orbit} and \ref{Section_waves} respectively.

	\section{Basic Energy Operations}\label{Section_basicEnergyOp}

	\begin{table}
		%\small
		\centering
		\caption{Examples of Energy Operations\label{EnOps}}
		\begin{tabular}{ll} 
			\hline
			Operation &  \textit{Identified where:} \\ \hline%\hline
			%\multicolumn{1}{c}{}
			%\rowcolor{Gray}
			{Method Invocation} & \textit{ one method is called}\\ %\hline
			
			Parameter\_Object &  \textit{ Object is one parameter of the method}\\ %\hline
			%&  Identified when``Object" is a method parameter \\	
			%\multirow{-2}{*}{Parameter\_Object} &  of the method \\
			
			Return\_Object &  \textit{ the method returns an Object}\\ %\hline
			
			%\cellcolor{Gray} & \cellcolor{Gray}Identified when symbol ``$+$" appears in code,  \\  
			%\multirow{-2}*{ \cellcolor{Gray} Addition\_int\_int}& \cellcolor{Gray}and two operands are integers\\ 
			%\rowcolor{Gray} 
			Addition\_int\_int & \textit{ addition's operands are integers }\\ %\hline

			Multi\_float\_float & \textit{ multiplication's operands are floats }\\
			%\hline
			%\rowcolor{Gray}
			Increment & \textit{ symbol "$++$" appears in code}\\ 
			%\hline
			And & \textit{ symbol "$\&\&$" appears in code}\\ 
			%\hline
			%\rowcolor{Gray}
			Less\_int\_float & \textit{ "<"'s operands are integer and float}\\%\hline
			%\cellcolor{Gray} & \cellcolor{Gray}Identified when symbol ``$<$" appears in code,  \\  
			%\multirow{-2}*{ \cellcolor{Gray} Less\_int\_float}& \cellcolor{Gray}and one operand is integer, %one is floating\\  
			
			Equal\_Object\_null & \textit{ "=="'s operands are Object and null}\\ %\hline
			%&  Identified when symbol ``$==$" appears in code, \\	
			%\multirow{-2}{*}{Equal\_Object\_null} &  and one operand is ``Object", one is ``null" \\
			
			%\rowcolor{Gray}
			Declaration\_int & \textit{ one integer is declared}\\%\hline
			
			Assign\_Object\_null & \textit{ assignment's operands are Object and null}\\
			%\hline
			%	&  Identified when symbol ``$=$" appears in code, \\	
			%	\multirow{-2}{*}{Assign\_Object\_null} &  and `null" is assigned to an ``Object" \\	
			%\rowcolor{Gray}	 
			Assign\_char[]\_char[] & \textit{ assignment's operands are arrays of chars}\\%\hline
			%\cellcolor{Gray} & \cellcolor{Gray}Identified when symbol ``$=$" appears in code,  \\  
			%\multirow{-2}*{ \cellcolor{Gray} Assign\_char[]\_char[]}& \cellcolor{Gray}and operands are two %arrays of chars \\  
			
			Array Reference &  \textit{ one array element is referred}\\
			%\hline
			%	\rowcolor{Gray}	 
			Block Goto & \textit{ the code execution goes to a new block}\\
			%\cellcolor{Gray} & \cellcolor{Gray}Identified when the code execution goes to  \\  
			%\multirow{-2}*{ \cellcolor{Gray} Block Goto}& \cellcolor{Gray}a new block \\ 

			\hline
		\end{tabular}
	\end{table}
	
	%If we use statements or source line as basic modeling units, 
	
	%The statement as a basic unit is not easy for modeling since any pair of statements are probably distinct. For instance, an arithmetic expression in the statement could have two operators or more, which could be either additions or multiplications. Alternatively, if we go to the method level, the model will be restricted to the target source code.
	
	Energy operations are identified directly from source code. The enumeration of the operations is inspired by Java semantics \cite{Bogdanas_Semantics}, which specifies the operational meaning, or behavior, of the Java language, which is the target language in the experiment.
	We intuitively identify semantic operations that perform operations on the state and may be energy-consuming, and let them be our energy operations. 
	%Also for a certain set of semantics, we extend the semantics to be the energy operations, for example, we append all the operands information to the 
	%as many energy-consuming operations as possible according to the Java semantics list,
	Ones that have little or no energy effect will automatically be identified by the regression analysis in the later stage of the analysis.
	Table \ref{EnOps} lists 14 representative operations out of a total of 120 in the experiment.
	They include arithmetic calculations like \textit{Multi\_float\_float}, \textit{Addition\_int\_int}, in which operands types are explicit, as well as \textit{Increment} whose operand is implicitly an integer.   Boolean operations and comparisons, such as \textit{And}, \textit{Less\_int\_float} and \textit{Equal\_Object\_null} also form one major part. \textit{Method Invocation} and \textit{Block Goto} are important for the control flow which plays a key role in the execution of the code. Assignments and \textit{Array Reference} will unexpectedly take a significant amount of the application's energy consumption, as will be shown in Section \ref{Section_Analysis}. 
	\begin{table}
		%\small
		\centering
		\caption{Examples of Library Functions\label{Libaray_functions}}
		\begin{tabular}{ll} \hline
			Class & $\qquad$Function \\ 
			\hline%\hline
			%\rowcolor{Gray}
			ArrayList & \textit{add, get, size, isEmpty, remove}   \\ %\hline
			
			& \textit{glBindTexture, glDisableClientState } \\
			& \textit{glDrawElements, glEnableClientState} \\
			GL10     & \textit{glMultMatrixf, glTexCoordPointer} \\
			& \textit{glPopMatrix, glPushMatrix} \\
			& \textit{glTexParameterx, glVertexPointer}\\
			%\hline
			%\rowcolor{Gray}
			Math    & max, pow, sqrt, random  \\%\hline
			
			FloatBuffer  & \textit{position, put} \\
			\hline
		\end{tabular}
	\end{table}
	
	The game engine application, like many others, also employs a diversity of library functions. Unlike the normal source code which is interpreted at run-time, the key part of library code has been compiled into native code before execution and some part may be already written in different languages and at lower levels of the software stack. On the other hand,  usually a limited number (67 in the experiment) of library functions are frequently called in one application. So we treat them as basic modeling units. The examples of highly-used library functions in the experiment are shown in Table \ref{Libaray_functions}. For instance, the functions in the class of \textit{GL10} are responsible for graphic computing.

	\section{Experimental Setup}\label{Section_experimentsetup}
	
	In this section and the next, we summarize the construction of the energy model, including the setup of the target device and the design principles of the execution cases.  Further details on these can be found in our recent work\footnote{To preserve anonymity in the review process, the citation is hidden}.
	
	\subsection{Target Device}\label{Section_target_measurement}
	Experimental target: we employ an Odroid-XU+E development board \cite{target:odroid} as the target device. It possesses two ARM quad-core CPUs, which are Cortex-A15 with 2.0 GHz clock rate and Cortex-A7 with 1.5 GHz. The eight cores are logically grouped into four pairs. Each pair consists of one big and one small core. So from the operating system's point of view there are four logic cores. In our experiment, we turn off the small cores and run workload on big cores at a fixed clock frequency of 1.1 GHz. We do this in order to remove the influence of voltage, clock rate and CPU performance on the power usage.   
	Odroid-XU+E has a built-in power monitoring tool to measure the voltage and current of CPUs with a frequency of 30 Hz. %and updates the samples in a log file. 
	
	%We wrote a script to obtain the samples from the file. During execution we run the script on an idle core to minimize its influence on the application.   
	
	%Note that the power monitor gives two sequences of power samples: one is for the big cores and the other is for the small cores. We pick the sequence of power samples of the big cores, because we only run workload on them.
	
	%the power reading has two segments: one is the power sample of four big cores, the other is the power sample of four small cores. We pick the big cores' power sample, because we only run workload on big cores.

	\subsection{Target Source Code}
	
	The target source code is the Cocos2d-Android \cite{code:cocos2d} game engine, a framework for building games, demos and other interactive applications such as virtual reality. It also implements a fully-featured physics engine. Games are increasingly popular on mobile phones and include more and more fancy and energy-consuming features, requiring high CPU performance. This paper demonstrates the energy modeling, accounting and improvement for the source code of the game engine, and evaluates the improvement in three game scenarios. 
	
	%but the methodology of energy modeling, accounting and code optimization is applicable for all kinds of applications.
	
	\subsection{Design of Execution Cases }\label{Section_sourcecode_casedesign}
	
	The execution cases whose energy usage is measured and analyzed represent typical sequences of actions during  game, including user inputs. We focus on three scenarios which are \texttt{Click \& Move}, \texttt{Orbit} and \texttt{Waves}.

	In the \texttt{Click \& Move} scenario, the sprite (the character in the game) moves to the position where the tap occurs. In the \texttt{Orbit} scenario, the sprite together with the grid background spins in the three-dimension space. In the \texttt{Waves} scenario, the sprite scales up and down, meanwhile the grid background waves like flow. In both the \texttt{Orbit} and \texttt{Waves} scenarios, the animation will restart from the starting point whenever and wherever the tap occurs. 
	
	To simulate the game scenarios under different sequences of user inputs, we script with the Android Debug Bridge \cite{adb:android} (ADB), a command line tool connecting the target device to the host, to automatically feed the input sequences to the target device.
	
	In order to obtain a more varied set of execution cases and thus a more precise model, we vary the executions of individual basic blocks in the code. This is achieved by systematically removing a set of blocks for each execution case, using the control flow graph extracted using the Soot tool \cite{soot:callgraph}. We ensure that each block could be removed in some execution case.
	%The problem is that a certain amount of blocks are critical to the functionality of the game engine, so we avoid removing them in the design stage.    
	Thus an execution case is made up of one user input sequence and one set of basic blocks.

\section{Model Construction}\label{Section:Model}

%The aimed model is formalized in Equation \ref{Energy_Model}. The entire energy consumption consists of the sum of the costs of operations and library functions and idle cost. Notice that the idle costs of individual cases are different, since they are executed in distinct sequences of inputs, and the lengths of sessions are also varying. So we measure the idle cost for individual cases.   
The entire energy use is composed of three parts: the cost of energy operations, the cost of library functions and the idle cost. The aimed model is formalized in Equation (\ref{Energy_Model}):

\begin{equation}\label{Energy_Model}
E = \sum_{}^{op_i \in Energy\,Ops} Cost_{op_i} \cdot  N_e(op_i) \qquad\qquad \end{equation}
\begin{displaymath}
+ \sum_{}^{func_i \in Lib\,Funcs} Cost_{func_i} \cdot  N_e(func_i) + Idle\;Cost
\end{displaymath}

 The cost of energy operations is the sum of $  Cost_{op_i} \cdot  N_e(op_i) $ (the cost of one operation multiplied by the number of its executions), where $op_i \in Energy\,Ops$; $Energy\,Ops$ is the set containing all the operations. The cost of library functions is the sum of $Cost_{func_i} \cdot  N_e(func_i)$ (the cost of one library function multiplied by the number of its executions), where $func_i \in Lib\,Funcs$; $Lib\,Funcs$ is the set of library functions.
The $Idle\; Cost$ is the energy consumption of the device when running no application, simply the Android system. The lengths of case sessions are diversified due to input sequences, so the $Idle\; Cost$ is different for each execution case.
The model construction is based on regression analysis, finding out the correlation between energy operations and their costs from the data obtained in the execution cases.

\section{The \texttt{Click \& Move} Scenario}\label{Section_clickmove}

In this section, we begin with energy accounting at operation and block level for the  \texttt{Click \& Move} scenario, after which we improve the most costly blocks focusing on the most expensive operations. We apply a similar approach in the other scenarios \texttt{Orbit} and \texttt{Waves} in Section \ref{Section_orbit} and Section \ref{Section_waves}; however for those cases we will only briefly summarize
energy accounting and focus on the code improvements. 

\subsection{Energy Accounting}\label{Section_Analysis}
The energy model of app source code based on energy operations facilitates comprehensive energy accounting from operation-level up to source-level. In this section, we will see the rank of the most expensive operations, and the contributions of different operations to the energy consumption of each block.  

\begin{figure}
	\centering
	\includegraphics[width = 0.44\textwidth]{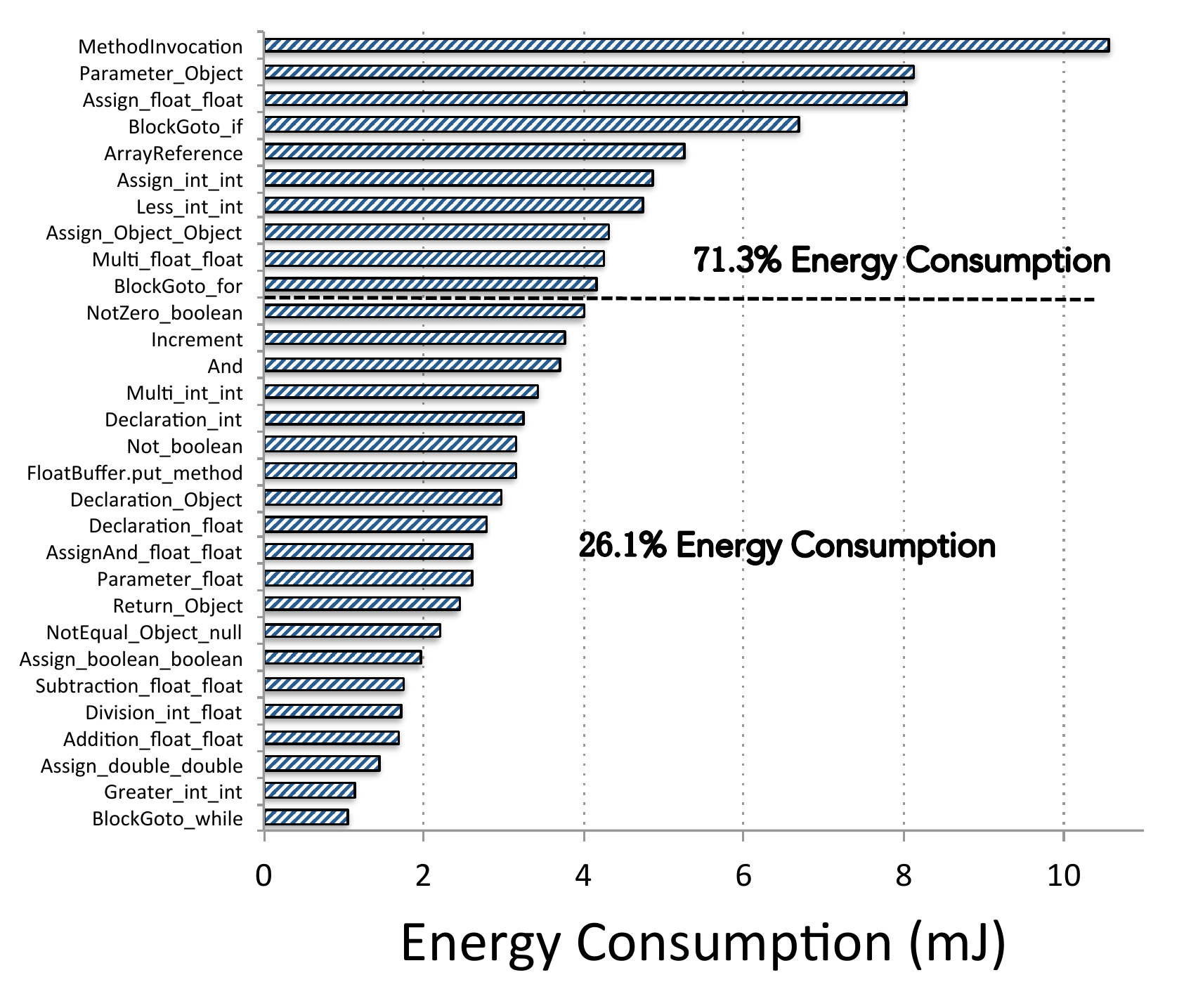}
	\caption{The top 30 energy consuming operations in \texttt{Click \& Move} scenario.}\label{fig:op_rank}
\end{figure}

\paragraph{Operation Level}
Figure \ref{fig:op_rank} shows the top 30 energy consuming operations in the model, ranked by their single-execution energy costs. The line marked
"71.3\% Energy Consumption" indicates the percentage of the energy cost of the execution cases for the \texttt{Click \& Move} scenario contributed by the top 10 operations. Similarly, "26.1\% Energy Consumption" shows the contribution to the total cost of operations from 11th to 30th. 
We can see that the energy usage of the code is largely determined (97.4\%) by a relatively small  number of operations. This is due to the fact that these operations are frequently used and expensive in themselves.
%The 30 operations out of 187 (including library functions) take up  of the total cost, in which the top 10 consumes the major part with a percentage of 71.3\%.
%Among all the operations (even beyond the top 30), the single-execution costs of them vary in a range of several orders of magnitude. 

It might be supposed that the sophisticated arithmetic operations, such as multiplications and divisions, should be the most costly. However, the result shows that \textit{Method Invocation} ranks the highest. This is due to a sequence of complex processes to fulfill \textit{Method Invocation}, for example, most of the method calls in Java are virtual invocations which are dispatched on the type of the object at run-time and always implicitly passed a "this" reference as their first parameter, not to mention other operations such as storing the return address and managing the stack frame.
%and \textit{Parameter\_Object}, two method-relevant operations,

This suggests a trade-off between code structure and energy saving when writing the code. That means, in certain cases, we could inline some thin and highly-invoked methods in the code, at the cost of losing the integrity of the structure of the code to some extent. 

Only one arithmetic operation, namely \textit{Multi\_float\_float}, is a member of the top 10, and there are only six arithmetic operations in the top 30. They together cost only 6.1\% of the overall energy consumption of the application, which is somewhat unexpected. 

%\textit{Increment}, \textit{Multi\_int\_int}, \textit{Subtraction\_float\_float}, \textit{Division\_int\_float} and \textit{Addition\_float\_float}. 
%Two arithmetic operations, \textit{Addition\_int\_int} and \textit{Multi\_float\_\\float}, are members of the top 10. Unexpectedly, the addition is twice as expensive as the multiplication. We surmise that this is a result of their operands in the target code, as experimentally shown in \cite{Steve_data_dependent}, the energy cost of the arithmetic computation is operand-dependent.

Later in block-level energy accounting, we will see that assignments, comparisons and \textit{Array Reference} play significant roles in the overall energy consumption. This is not only because they are frequently used, but also because they are costly as operations themselves, as shown in Figure \ref{fig:op_rank}. 

\textit{Block Goto} operations are expensive as well.
Based on the types of conditionals and loops where "Block Goto" occurs, they are classified into \textit{BlockGoto\_if}, \textit{BlockGoto\_for} and \textit{BlockGoto\_while}. The result shows that they cost different amounts of energy as operations themselves, respectively 6.7 $\mu$J, 4.1 $\mu$J and 1.1 $\mu$J. Together with \textit{Method Invocation}, they take up 37.6\% of the total energy consumption of the application. 

%\subsection{Block Level}\label{Section_blocklevelenergy}
\paragraph{Block Level}

\begin{figure*}
	\centering
	\begin{subfigure}[b]{0.49\textwidth}
		\centering            
		\includegraphics[width = \textwidth]{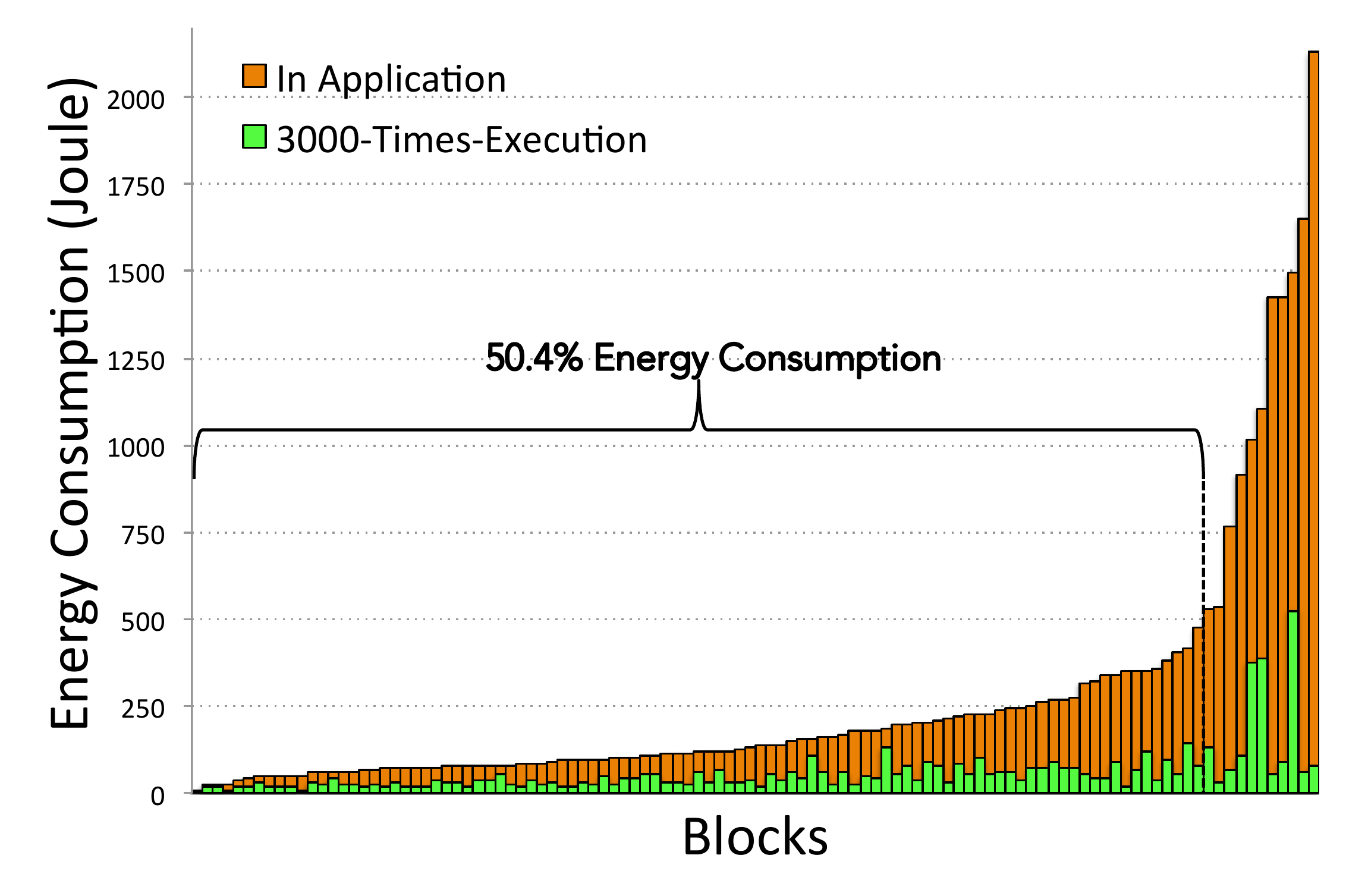}
		%\caption{ Block costs In Application and at 3000 Times Execution.}
		\label{fig:blocks_in_program}
	\end{subfigure}
	%add desired spacing between images, e. g. ~, \quad, \qquad etc.
	\begin{subfigure}[b]{0.49\textwidth}
		\centering
		\includegraphics[width=\textwidth]{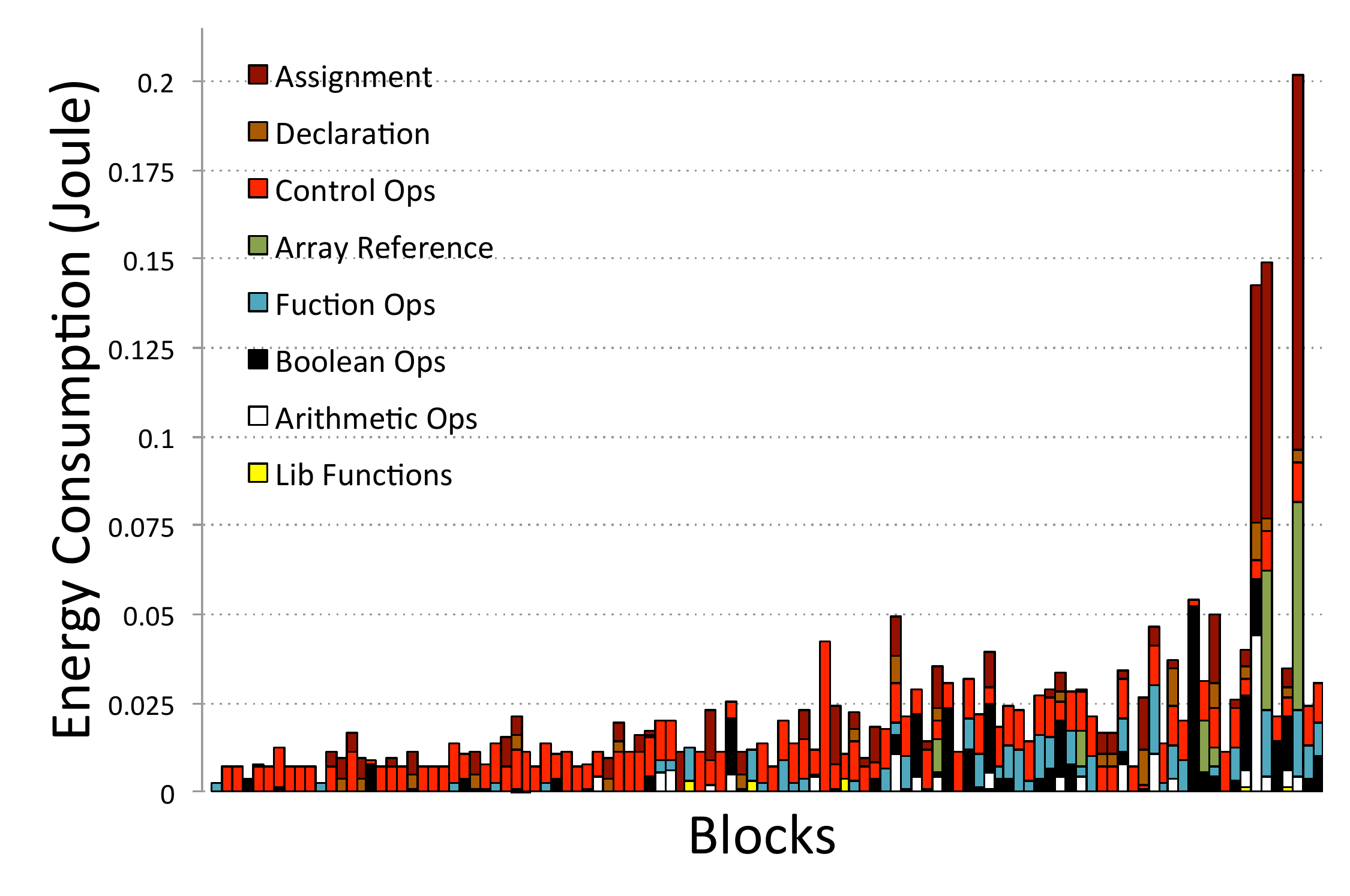}
		%\caption{Energy proportions of different kinds of operations in blocks.}
		\label{fig:op_in_block}
	\end{subfigure}
	\caption{Energy distribution in \texttt{Click \& Move}. Blocks are sorted by the order of their run-time energy costs In Application.}\label{fig:block}
\end{figure*}

In the execution cases, we have 108 active blocks with a wide diversity of energy usage. In Figure \ref{fig:blocks_in_program}, "In Application" means running the \texttt{Click \& Move} scenario with the full set of blocks (that is, ignoring the execution cases described in Section \ref{Section_sourcecode_casedesign} in which some blocks are removed). The total cost of a block "In Application" is plotted as an orange bar, reflecting both its cost and the number of times it is executed.
%Note that blocks here obviously have distinct execution times. 
The cost of a fixed number (3000) of executions of one block are calculated by multiplying its single-execution cost by 3000. This helps us to compare the single-execution costs of different blocks. The costs of blocks at "3000-Times-Execution" are plotted as green bars.

Similar to energy distribution on operations, only a small number (11 blocks) of all the blocks use up nearly half of the entire cost, which indicates that putting efforts on optimising a small group of blocks can achieve significant energy saving.

There are two factors that make one block costly "In Application". The first factor is a large number of executions. For example, the most costly block  "In Application" (the rightmost orange bar in Figure \ref{fig:blocks_in_program}) has a large number of execution times. This block takes only 30.6 $\mu$J for single-execution but 2.1 Joules when running "In Application". 
The second factor is the energy consumption of the block itself. For example, we can see three prominent green bars in Figure \ref{fig:blocks_in_program}, whose single-execution costs are 201.5 $\mu$J, 146.9 $\mu$J and 142.8 $\mu$J. We will later zoom in these three blocks to see which operations contribute to their energy costs.

%the contributions of different kinds of operations.

We can further observe the energy proportions of operations in each block in Figure \ref{fig:op_in_block}. To illustrate, operations are grouped into eight classes. Specifically, the "Block Goto" operations and \textit{ Method Invocation} are gathered in \textit{Control Ops}; the parameter passing and the value returns of methods are in \textit{Function Ops}; the comparisons and Booleans are in \textit{Boolean Ops}; all the arithmetic computations are in \textit{Arithmetic Ops}; all the library functions are in \textit{Lib Functions}.

Most of the blocks cost less than 25 $\mu$J for single-execution. In these blocks, \textit{Control Ops} occupy the major part of the energy consumption, in contrast, \textit{Arithmetic Ops} only take a tiny proportion. 

For those three most prominent blocks, assignments and \textit{Array Reference} are the biggest energy consumers. Furthermore one of the three blocks has the largest proportion of \textit{Arithmetic Ops} among all the blocks.  

The most expensive block "In Application" consists of three even parts: \textit{Control Ops}, \textit{Function Ops} and \textit{Boolean Ops}. This block is the main entrance of the game engine to draw and display frames, so its work is dominated by conditional judgments and method invocations.  

\subsection{Code Optimization}
\begin{table}
	%\small
	\centering
	\caption{The top 10 most costly blocks in \texttt{Click \& Move}. \label{table_topblocks}}
	\begin{tabular}{lr} \hline
		\quad Block ID  & Energy Cost (mJ)\\ 
		\hline%\hline
		CCNode.visit()  & 2128.6 \qquad\quad \\
		CCNode.transform() & 1648.4 \qquad\quad\\
		CCTextureAtlas.putVertex()  & 1494.4 \qquad\quad\\
		CCNode.visit().if\_4.for\_1 & 1426.8 \qquad\quad\\
		CCNode.transform().if\_1    & 1426.3 \qquad\quad\\
		CCTextureAtlas.putTexCoords() &  1107.8 \qquad\quad\\
		CCAtlas.updateValues().for\_1 & 1018.7 \qquad\quad\\
		CCNode.visit().if\_3.for\_1  & 915.7 \qquad\quad\\
		CCSprite.draw() & 766.9 \qquad\quad\\
		CCTexture2D.name() & 537.5 \qquad\quad\\
		\hline
	\end{tabular}
\end{table}

The most important consideration of app developers is to guarantee the correctness of software, which should then be followed by energy efficiency. So our energy-aware programming approach is adopted at the end of software engineering life circle when the software system is in general complete. 

%The overview of energy-aware programming approach is firstly finding the most costly blocks, where we analyze the energy breakdown on the operations, and make changes to the code to diminish the usage of costly operations.

We look into the top 10 costly blocks "In Application" (see Table \ref{table_topblocks}). For example, \textit{CCNode.visit()} is the entrance block of the \textit{visit()} function; \textit{CCNode.visit().if\_4.for\_1} is the body block of the \texttt{for} loop.
These 10 blocks are distributed in seven methods, so the code review is straightforward. We find four easy opportunities to improve energy efficiency of some blocks:  \textit{CCNode.visit()}, \textit{CCNode.visit().if\_4.for\_1} and \textit{CCTexture2D.name()}. There are also other opportunities in other blocks supposed possible to save energy, but requiring more efforts and gaining little. For example, \textit{CCAtlas.\\updateValues().for\_1} has several busy arithmetic expressions. Usually it is supposed that replacing the busy expression with a variable would reduce energy, however in this case the overhead of variable declaration counteracts the saved energy. 

The four opportunities to reform the code are very simple and effective, but can only be discovered by the operation-level information. The changes are shown as follows:

\begin{program}
	\begin{lstlisting}
	if (children_ != null) {
	if_body1;
	}
	draw(gl);
	if (children_ != null) {
	if_body2;
	}
	
	\end{lstlisting}
	\caption{ Simplified parts of \textbf{original} code in \textit{CCNode.visit()}}
	\label{program1}
\end{program}

\begin{program}
	\begin{lstlisting}
	if (children_ != null) {
	if_body1;
	draw(gl);
	if_body2;
	} else {draw(gl);}
	\end{lstlisting}
	\caption{The changed Program \ref{program1} }
	\label{program2}
\end{program}

\paragraph{If Combination}

This change is made in the most costly block \textit{CCNode.visit()}, which has two comparisons, two Boolean operations, one \textit{Method Invocation} and one parameter passing. In fact, the two \texttt{if} headers make the same comparison, as shown in Program \ref{program1}. We change the code to Program \ref{program2}, which combines the two \texttt{if} statements and meanwhile keep it logically consistent with Program \ref{program1}. By these means each execution of the block can reduce one comparison, and when the condition is false, it can additionally reduce one \textit{BlockGoto\_if}. 

\begin{program}
	\begin{lstlisting}
	public void visit(GL10 gl) {
	......
	transform(gl);
	......
	}
	public void transform(GL10 gl) {
	tranform_body;
	}
	\end{lstlisting}
	\caption{Simplified parts of \textbf{original} code in \textit{CCNode} class}
	\label{program3}
\end{program}

\begin{program}
	\begin{lstlisting}
	public void visit(GL10 gl) {
	......
	transform_body;
	......
	}
	public void transform(GL10 gl) {
	transform_body;
	}
	\end{lstlisting}
	\caption{The changed Program \ref{program3}}
	\label{program4}
\end{program}

\paragraph{Inner-Class Method Inline}

When "In Application", the \textit{transform()} function is invoked 18903 times and mostly by the\textit{visit()} function.
We change the Program \ref{program3} to Program \ref{program4} by switching the body of \textit{transform()} to the function call of \textit{transform()} in \textit{visit()}, meanwhile remaining the original definition of \textit{transform()} in case that other parts of the code call it. This change can greatly decrease the number of calls to \textit{transform()}s and thus \textit{Method Invocation}s that are costly. However, it may be at the cost of losing readability of the code (which might be partly compensated by adding explanatory comments). 

\paragraph{Loop-Invariant Code Motion}

\textit{CCNode.visit().if\_3.for\_1} and \textit{CCNode.visit().if\_4.for\_1} are entrance blocks of the two \texttt{for} loops as seen in Program \ref{program5}. These two loops share a quantity, \textit{children\_.size()}, which is computed in each iteration but actually constant. We thus hoist it outside the loop, as shown in Program \ref{program6}, which vastly saves the energy of invoking and executing the \textit{size()} function during every iteration. Meantime, we move the declaration of the \textit{child} outside the loop, considering the cost of \textit{Declaration\_Object} is about 2.97 $\mu$J and also in the top 30. \\

\paragraph{Inter-Class Method Inline}

\textit{CCTexture2D.name()} is the 10th most costly block and costs 537.5 mJ "In Application". However, its job is to simply get the value of the private member variable, \textit{\_name}, of the class \textit{CCTexture2D}. This method has only two callers in the code. So we consider to make this variable public and let the two callers directly get access to the variable, which avoids the cost of \textit{Method Invocation}. This change may harm the encapsulation of data, however, only one member of one class is changed. The trade-off between energy-saving and data encapsulation will be at last decided by developers. 

\begin{program}
	\begin{lstlisting}
	if (children_ != null) {
	for (int i=0; i<children_.size(); ++i) { 
	CCNode child = children_.get(i);
	if (child.zOrder_ < 0) {
	child.visit(gl);
	} else
	break;
	}
	draw(gl);
	for (int i=0; i<children_.size(); ++i) { 
	CCNode child = children_.get(i);
	if (child.zOrder_ >= 0) {
	child.visit(gl);
	}
	}
	} else {draw(gl);}
	\end{lstlisting}
	\caption{The full version of Program 2}
	\label{program5}
\end{program}

\begin{program}
	\begin{lstlisting}
	CCNode child = new CCNode(); //added
	int children_size = children_.size(); //added
	if (children_ != null) {
	for (int i=0; i<children_size; ++i) { //changed 
	child = children_.get(i); //changed
	if (child.zOrder_ < 0) {
	child.visit(gl);
	} else
	break;
	}
	draw(gl);
	for (int i=0; i<children_size; ++i) { //changed
	child = children_.get(i); //changed
	if (child.zOrder_ >= 0) {
	child.visit(gl);
	}
	}
	} else {draw(gl);}
	\end{lstlisting}
	\caption{The changed Program \ref{program5}}
	\label{program6}
\end{program}

\subsection{Evaluation}

\begin{figure}
	\centering
	\includegraphics[width = 0.45\textwidth]{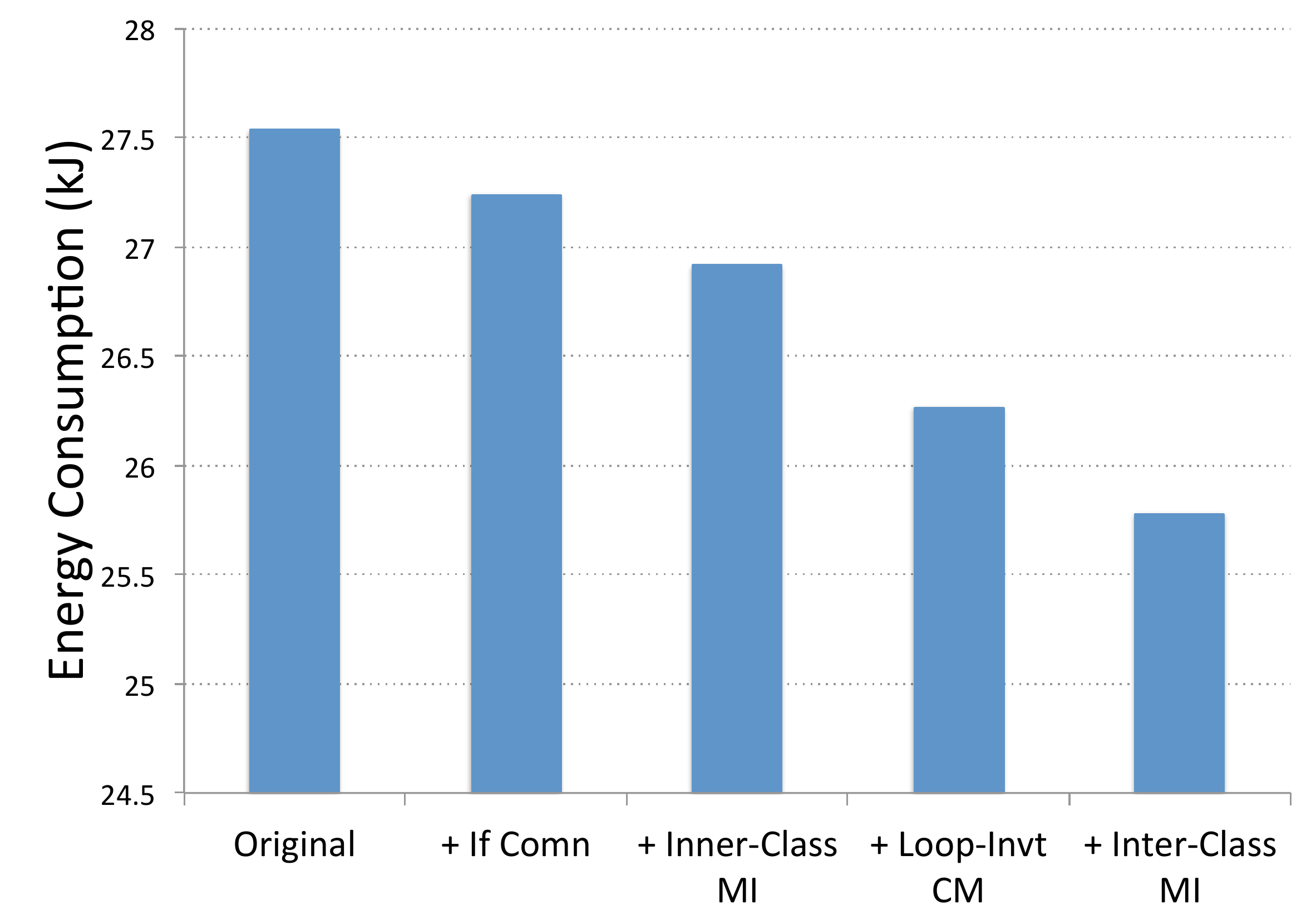}
	\caption{Energy consumption of the code without and with the changes in \texttt{Click \& Move}.}\label{fig:energy_saving_candm}
\end{figure}

Figure \ref{fig:energy_saving_candm} illustrates the energy dissipation of the software without and with the changes introduced in the previous section. From left to right, the bars indicate cumulative effects of the changes. For example, "\textit{+ If Comn}" is the energy consumption of the original code with the change of "If Combination"; "\textit{+ Inner-Class MI}" is the energy consumption of the code with the changes of both "If Combination" and "Inner-Class Method Inline". In total, these four simple changes save 6.4\% of the entire energy consumption without influencing the functionality of code. These changes are made in the basic part of the game engine, which most applications will be bases on, so any gain here can have fundamental impact. Furthermore, these changes are made with little knowledge about the algorithm of the code, the developers who designed the code are surely able to improve the code much more and achieve far more energy-saving, if the energy model was available to them.

\section{The \texttt{Orbit} Scenario }\label{Section_orbit}

In this section, we briefly describe the energy accounting for the \texttt{Orbit} scenario. Then we improve the most costly blocks focusing on the expensive operations. In Section \ref{orbit_evaluation}, the experimental result shows that the improvement can save 50.2\% of the overall energy consumption. 

\subsection{Energy Accounting}

\begin{figure}
	\centering
	\includegraphics[width = 0.38\textwidth]{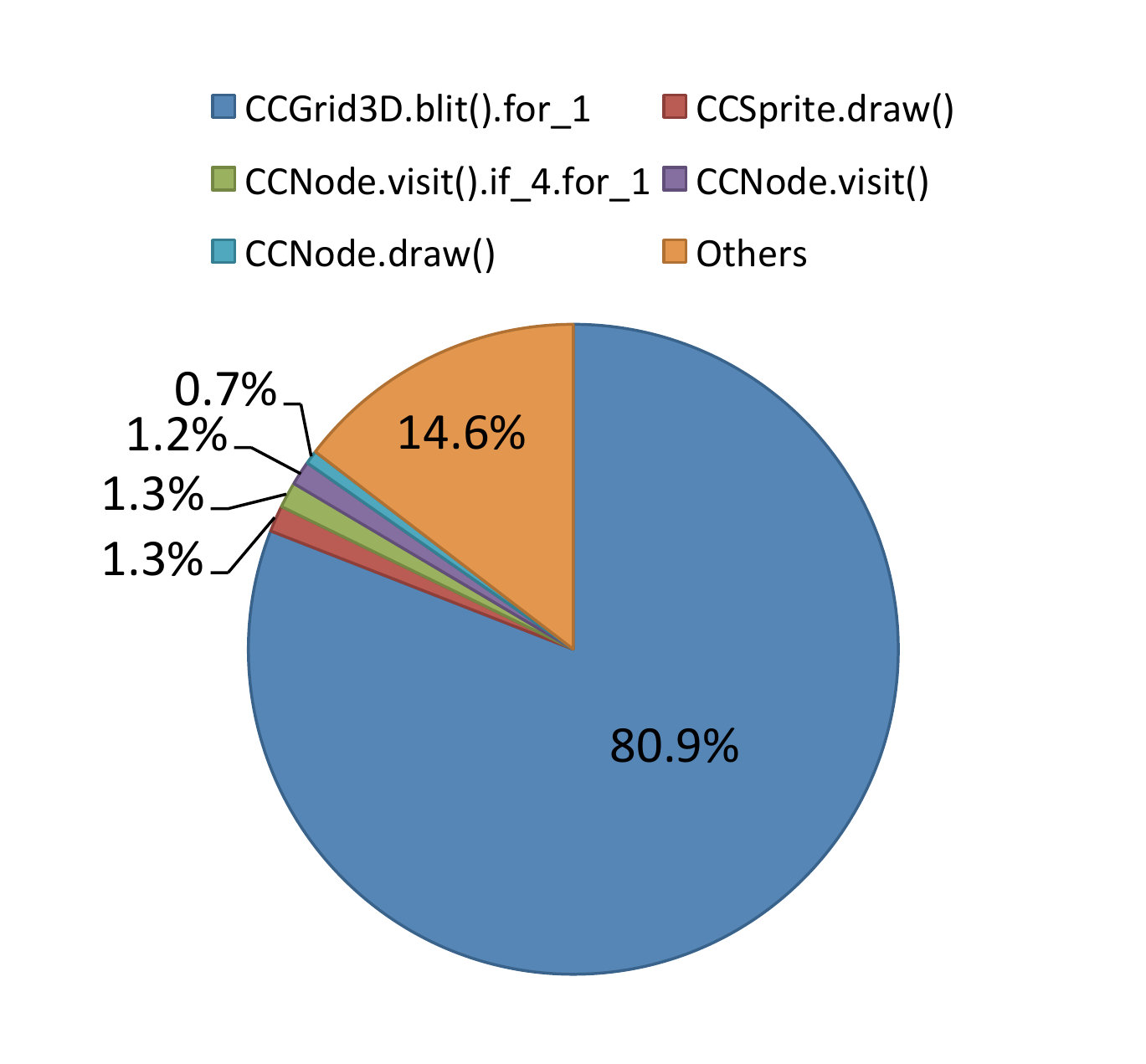}
	\caption{The energy proportions of blocks "In Application" in the \texttt{Orbit} scenario}\label{fig:lensorbit_blocks}
\end{figure}

In the \texttt{Orbit} scenario, the block \textit{CCGrid3d.blit().for\_1} dominates the overall energy consumption. As shown in Figure \ref{fig:lensorbit_blocks}, 80.9\% of the entire cost is consumed by this block. The second most costly block consumes only 1.3\%. "In Application" here means running the \texttt{Orbit} scenario without removing any block. Later in Section \ref{section:orbit_code_optn}, we only focus on this single block. 
%It will be seen that a little effort is made to achieve eminent improvements.   

\subsection{Code Optimization}\label{section:orbit_code_optn}

Program \ref{program7} shows the original code of \textit{CCGrid3D.blit().\\for\_1}. In this block, the \textit{Control Ops} (\textit{BlockGoto\_for} and \textit{Field Reference}) use up 35.6\% of the energy; \textit{Boolean Ops} use up 20.5\%; the assignments use up 16.7\%; \textit{Arithmetic Ops} use up 14.0\%; \textit{Lib Functions} use up 13.3\%. We find three easy changes to reduce or replace the pricey operations.\\

\paragraph{Loop-Invariant Code Motion}

In this block, the value of \textit{vertices.limit()} is the constant 2112; we therefore hoist it outside the loop and replace it with the variable \textit{limit}, as shown in Program \ref{program8}. This change avoids invocations and executions of \textit{vertices.limit()} and at the same time decreases a small amount of \textit{Field Reference}. 

\begin{program}
	\begin{lstlisting}
	for (int i = 0; i < vertices.limit(); i=i+3) {
	mVertexBuffer.put(vertices.get(i)); 
	mVertexBuffer.put(vertices.get(i+1));
	mVertexBuffer.put(vertices.get(i+2));
	}
	\end{lstlisting}
	\caption{The \textbf{original} code of \textit{CCGrid3D.blit().for\_1}}
	\label{program7}
\end{program}

\begin{program}
	\begin{lstlisting}
	int limit = vertices.limit(); //added
	for (int i = 0; i < limit; i=i+24) { //changed
	mVertexBuffer.put(vertices.get(i)); 
	mVertexBuffer.put(vertices.get(i+1));
	mVertexBuffer.put(vertices.get(i+2));
	...                       
	mVertexBuffer.put(vertices.get(i+23));//added
	}
	\end{lstlisting}
	\caption{The changed Program \ref{program7}}
	\label{program8}
\end{program}

\paragraph{Loop Unrolling}
Also as shown in Program \ref{program8}, we duplicate the loop body eight times, reducing the times of comparisons, \textit{BlockGoto\_for}s, assignments and additions. Note that we set the value of the increment as 24 since 24 is a factor of the \textit{limit}, 2112.  

\paragraph{Full Use of Library Function}

The job of Program \ref{program7} or Program \ref{program8} is to get all the elements in \textit{vertices} one by one and put them one by one into \textit{mVertexBuffer}.  Program \ref{program7} can be simply replaced by one line: \textit{mVertexBuffer.put(vertices.asReadOnlyBuffer())}. This puts all the elements of \textit{vertices} into \textit{mVertexBuffer}. This change realizes the same functionality using the already existing library function, which is one of the key library functions already compiled into native code. 

\subsection{Evaluation}\label{orbit_evaluation}

\begin{figure}
	\centering
	\includegraphics[width = 0.41\textwidth]{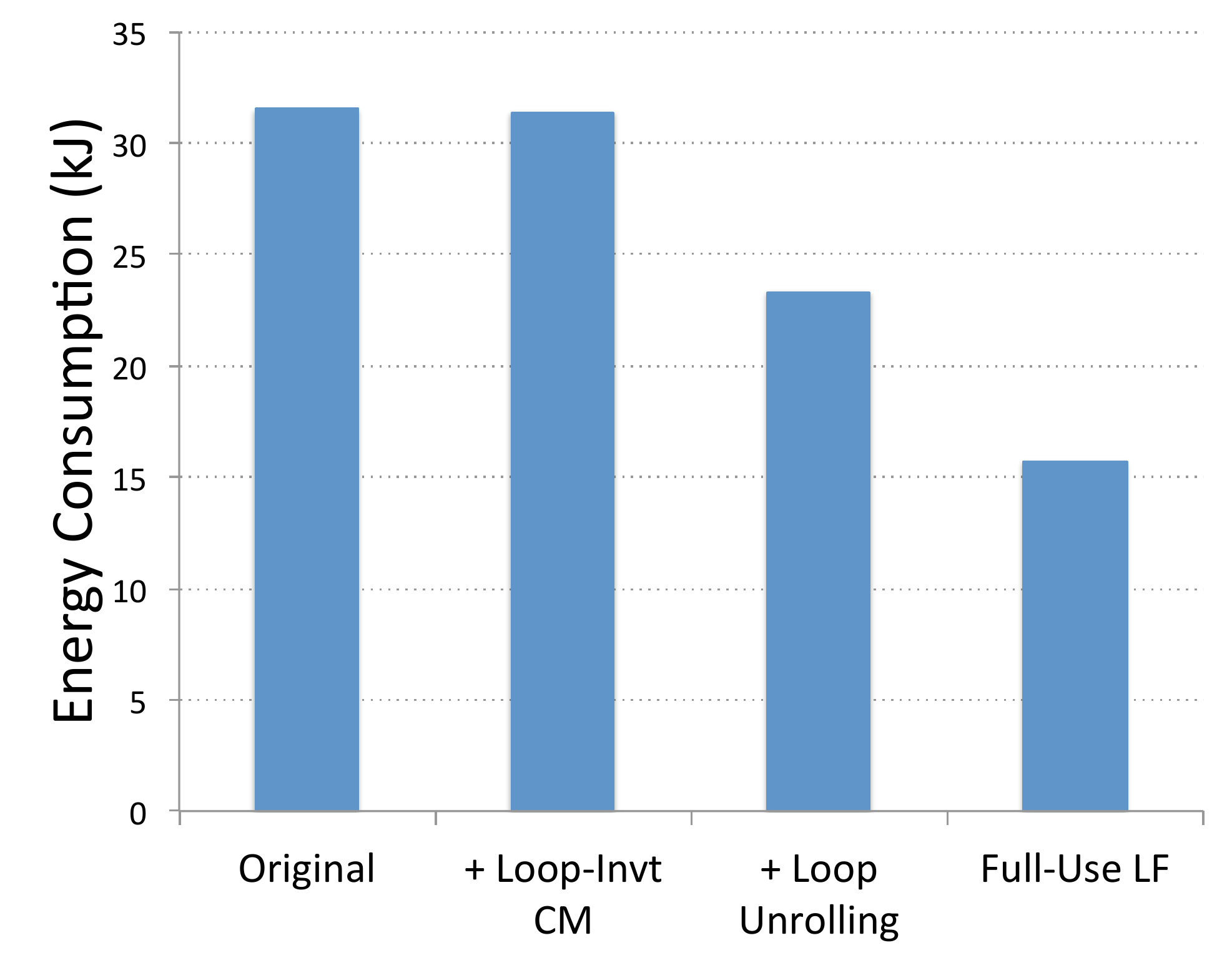}
	\caption{Energy consumption of the code without and with the changes in \texttt{Orbit}.}\label{fig:energy_saving_orbit}
\end{figure}

Figure \ref{fig:energy_saving_orbit} shows the cumulative effects of the code changes on energy consumption. In contrast to the other columns, "\textit{Full-Use LF}" does not take previous changes into account and means only replacing Program \ref{program7} with the built-in library function as stated above. The figure shows that loop-invariant code motion does not gain much energy saving because  \textit{vertices.limit()} is a library function and in addition uses a very small percentage of energy consumption. On the other hand, loop unrolling achieves 25.8\% energy saving due to the reduction of the amount of \textit{Control Ops}, comparisons and assignments, which occupy most of the cost. The most effective change is the replacement to the library function, avoiding the waste of 50.2\% energy use because 
this library function has been compiled into native code before execution, in contrast the Java source code need run-time interpretation which of course incurs an energy cost. The result implies that it is a good idea for developers to make a good use of library functions rather than implementing the same function with Java source code.  The discovery of this source of inefficiency was assisted by the energy accounting.

\begin{table*}
	%\small
	\centering
	\caption{Top 10 most costly blocks "In Application" in the \texttt{Waves} scenario and the energy percentages of different kinds of operations in each block.}\label{table_waves_topblocks}
	\begin{tabular}{lrrrrrrrrr} \hline
		\quad Block ID & \#Executions & Energy Cost (mJ) & Assi. & Decl. & Cont. & Func. & Bool. & Arit. & Libr. \\ 
		\hline%\hline
		CCGrid3D.blit().for\_1 & 112193 \quad\quad &  8094.1 \qquad\quad & 16.7\% & 0\% & 35.6\% &  0\% & 20.5\% &  14.0\% & 13.3\% \\
		CCVertex3D.CCVertex3D() & 40219 \quad\quad & 5232.0 \qquad\quad & 27.2\% & 0\% & 10.0\% & 62.8\% & 0\% &  0\% & 0\%\\
		CCWaves3D.update().for\_1.for\_1 & 34604 \quad\quad& 4088.7 \qquad\quad & 10.7\% & 0\% & 32.1\% & 0\% & 14.7\% &  39.0\% & 2.2\%\\
		ccGridSize.ccg() & 42275 \quad\quad& 3769.1 \qquad\quad & 0\% & 0\% & 32.1\% & 67.9\% & 0\% &  0\% & 0\%\\
		CCGrid3DAction.setVertex()   & 31856 \quad\quad & 3285.4 \qquad\quad & 14.6\% & 7.8\% & 30.9\%  & 46.7\% & 0\% &  0\% & 0\%\\
		CCGrid3DAction.originalVertex() & 36566 \quad\quad &  2891.3 \qquad\quad & 19.1\% & 10.2\% & 40.3\% & 30.4\% & 0\% &  0\% & 0\%\\
		CCNode.getGrid() & 49119 \quad\quad& 2145.1 \qquad\quad & 0\% & 0\% & 58.1\% & 41.9\% & 0\% &  0\% & 0\%\\
		ccGridSize.ccGridSize() & 10570 \quad\quad & 1173.8 \qquad\quad & 30.3\% & 0\% & 31.6\% & 38.0\% & 0\% &  0\% & 0\%\\
		CCGrid3D.setVertex() & 3944 \quad\quad& 657.2 \qquad\quad & 10.1\% & 1.6\% & 32.8\% & 28.9\% & 0\% &  26.4\% & 0.2\%\\
		CCGrid3D.originalVertex() & 2785 \quad\quad& 374.2 \qquad\quad & 14.0\% & 1.9\% & 33.4\% & 17.9\% & 0\% &  32.8\% & 0\%\\ 
		\hline
	\end{tabular}
\end{table*}

\section{The \texttt{Waves} Scenario}\label{Section_waves}

In this section, similarly, we first analyze the energy features of the blocks in the \texttt{Waves} scenario, based on which we modify the code and then evaluate the effects of changes on energy consumption.  

\subsection{Energy Accounting}

Unlike the \texttt{Orbit} scenario where only one block dominates energy cost, in the \texttt{Waves} scenario the costs of the top eight blocks are at the same order of magnitude of kJ, as listed in Table \ref{table_waves_topblocks}. The \textit{CCGrid3D.blit().for\_1} is also employed in this scenario and is the most costly as well among all the blocks. The majority of blocks in Table \ref{table_waves_topblocks} are directly or indirectly invoked by \textit{CCWaves3D.update().for\_1.\\for\_1}, as shown in Program \ref{program9}.   The purpose of these methods is mainly to set or get the values of member variables, so a large part of energy consumption goes to assignments, \textit{Function Ops} and \textit{Control Ops}. It was not expected that the code spends such a large amount of energy on simple set and get functions. 

\subsection{Code Optimization}

\paragraph{Full-Use of Library Function}

We mentioned previously in Section \ref{section:orbit_code_optn}  the optimization for \textit{CCGrid3D.blit().for\_1} where we replace the entire Program \ref{program7} with one line of code making use of library functions. We keep this change in this scenario. For other blocks, we come up with one modification as below.\\

\begin{program}
	\begin{lstlisting}
	int i, j;
	for( i = 0; i < (gridSize.x+1); i++ ) {
	for( j = 0; j <(gridSize.y+1); j++ ) {
	CCVertex3D v=originalVertex(ccGridSize.ccg(i,j));
	...
	setVertex(ccGridSize.ccg(i,j), v);
	}
	}
	\end{lstlisting}
	\caption{The \textbf{original} code in \textit{CCWaves3D.update()}}
	\label{program9}
\end{program}
\normalsize

\begin{program}
	\begin{lstlisting}
	ccGridSize ccgridsize = new ccGridSize(0,0);//added
	CCGrid3D ccgrid3d = 
	(CCGrid3D) target.getGrid(); //added
	CCVertex3D	v = new CCVertex3D(0,0,0); //added
	int i, j;
	for( i = 0; i < (gridSize.x+1); i++ ) {
	for( j = 0; j <(gridSize.y+1); j++ ) {
	ccgridsize.x=i;ccgridsize.y=j;  //added
	v =ccgrid3d.originalVertex(ccgridsize);//changed
	...
	ccgrid3d.setVertex(ccgridsize, v);//changed  
	}
	}
	\end{lstlisting}
	\caption{Program \ref{program9} after Method Inline \& Code Motion   }
	\label{program10}
\end{program}

\iffalse
\begin{program}
	\begin{lstlisting}
	...
	for( i = 0; i < (gridSize.x+1); i++ ) {
	ccgridsize.x=i;ccgridsize.y=0;   
	...
	ccgrid3d.setVertex(ccgridsize, v);
	ccgridsize.x=i;ccgridsize.y=1; //added  
	...
	ccgrid3d.setVertex(ccgridsize, v); //added
	...
	...
	ccgridsize.x=i;ccgridsize.y=10; //added  
	...
	ccgrid3d.setVertex(ccgridsize, v); //added         
	}
	\end{lstlisting}
	\caption{Program \ref{program10} after Loop Unrolling }
	\label{program11}
\end{program}
\fi

\paragraph{Method Inline \& Code Motion}

As shown in Program \ref{program9}, the three functions called in the inner loop body are \textit{CCGrid3DAction.originalVertex()}, \textit{ccGridSize.ccg()} and \textit{CCGrid3DAction.setVertex()}, which respectively cost 2891.3 mJ, 3769.1 mJ and 3285.4 mJ "In Application". Note that, \textit{CCGrid3DAction} is the parent class of \textit{CCWaves3D}, so in Program \ref{program9} \textit{originalVertext()} and \textit{setVertex()} can be directly called without referring to their class names. 
As seen in Program \ref{program10}, we unpack these three methods in this block: the first and fourth "added" lines are unpacked \textit{ccGridSize.ccg()}; the second "added" and first "changed" lines are unpacked \textit{CCGrid3DAction.originalVertex()}; the second "added" and second "changed" lines are unpacked \textit{CCGrid3DAction.setVertex()}. This change removes all the \textit{Method Invocation}s, parameter passing and value returns related to these three functions invoked by this block. Note that the first three "added" lines are located outside the loop in order to reduce energy consumption of the process of initializing objects and calling \textit{CCNode.getGrid()}. 

%\textbf{Loop Unrolling: }

%As shown in Program \ref{program11}, we multiply the loop body 11 times to remove the inner loop since the value of \textit{gridSize.y} is constantly 10. This modification reduces \textit{BlockGoto\_for} operations, increments and comparisons. 

\subsection{Evaluation}

\begin{figure}
	\centering
	\includegraphics[width = 0.38\textwidth]{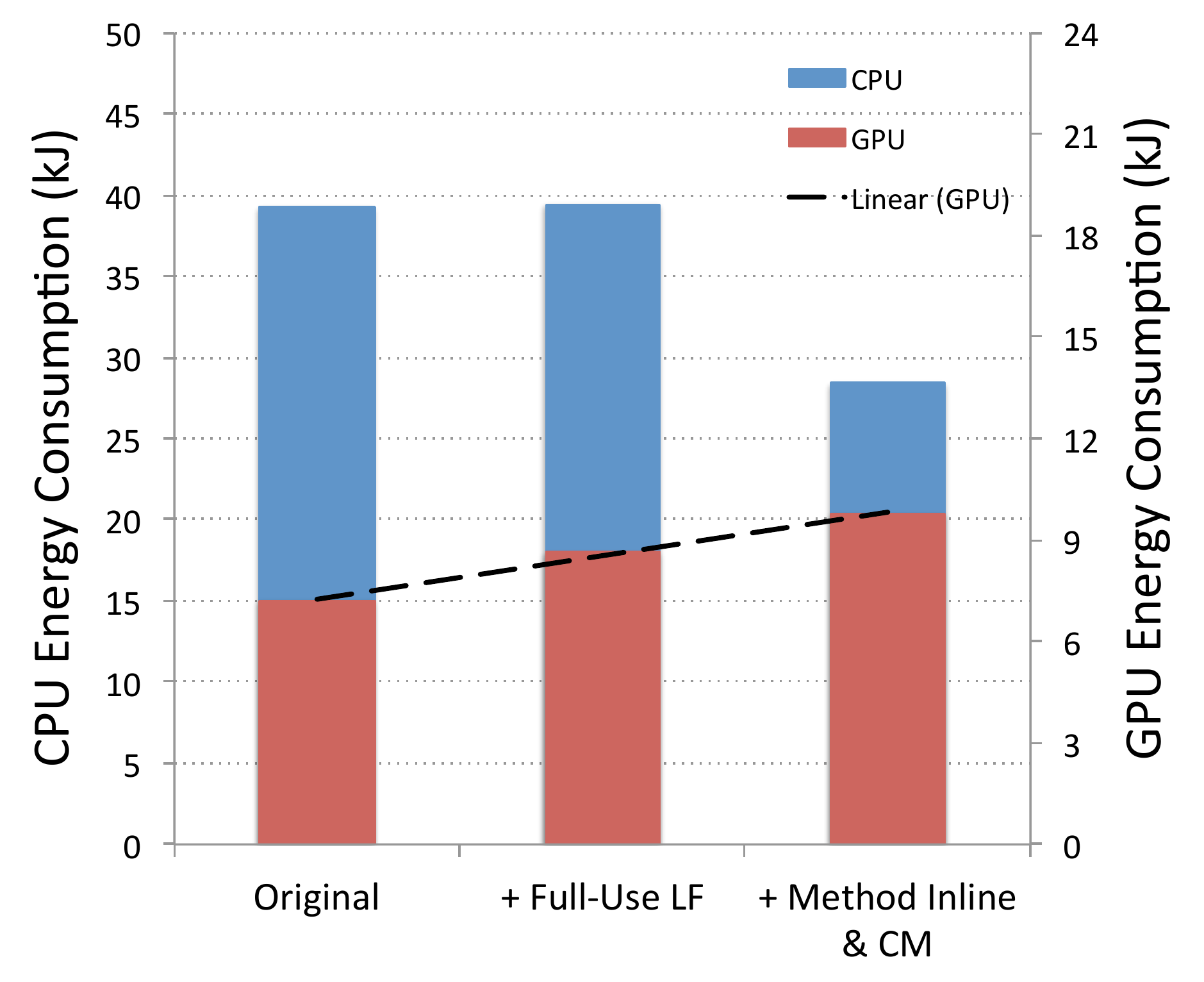}
	\caption{CPU and GPU Energy consumption of the code without and with the changes in \texttt{Waves}.}\label{fig:energy_saving_waves}
\end{figure}

Figure \ref{fig:energy_saving_waves} shows the cumulative effects of changes on energy consumption of CPU and GPU (note that previous figures only showed the CPU energy consumption because the GPU energy consumption did not vary noticeably), and the dashed line indicates the linear trend of the GPU energy consumption.
In the case of games,  the target frame rate is usually 60 Hz; when the game overloads the CPU the rate will decrease, and when the workload is light, even very light, the rate is generally fixed to 60 Hz. The rate in "\textit{Original}" is around 36 Hz; that in "\textit{+ Full-Use LF}" is around 50 Hz; that in "\textit{+ Method Inline \& CM}" is around 60 Hz.
The change of \textit{Full-Use LF} (full use of library function) does not save energy in the CPU because the execution of the original \texttt{Waves} actually overloads the CPU capacity, so the improvement of code enables the device to generate more frames every second. Consequently, the CPU does the same volume of work and consumes the equal amount of energy, the GPU does more work and consumes more energy, as seen in Figure \ref{fig:energy_saving_waves}. After this change, when we apply the method inline and code motion, 27.7\% of the overall CPU energy is saved, and for the same reason the GPU consumes slightly more. This experimental result indicates that our approach not only saves energy but also potentially boosts performance, which benefits the user doubly.

%At this moment, the device is capable to fulfill the performance requirement of the improved code, after which when we apply the change of method inline and code motion, 27.7\% of the overall CPU energy is saved. This experimental result shows that our approach not only saves energy but also boosts performance, which benefits users doubly.    

\section{Related Work}

\paragraph{Energy Modeling}

From the hardware side, research on energy modeling have been done at the circuit level (see the survey \cite{Najm_VLSI_level}), gate level \cite{Najm_gate_level, Marcu_gate_level} and register-transfer level \cite{cheng_RT_level}. Later, research focus shifted towards high-level modelings, such as software and behavioral levels \cite{Macii_high_level}.

Energy modeling techniques for software start with the basic instruction level, which calculates the sum of energy consumption of basic instructions and transition overheads \cite{Tiwari:power_analysis_embedded,bran:instruction-level_model}.   
Gang et al. \cite{gangqu:function-level_powermodel} base the model at the function-level while considering the effects of cache misses and pipeline stalls on functions. T. K. Tan et al. \cite{Tan:2001:high-level_softwaremodel} utilize regression analysis for high-level software energy modeling. 

However, the run-time context considered in the above works is unsophisticated, free from user inputs, a virtual machine, dynamic compilation, and etc. Furthermore the software stack below the level that they deal with (such as the level of the basic or assembly instruction) is relatively thin. 

When research is focused on the energy use of mobile applications, the level of granularity of the techniques is increased as well. An important part of such efforts is the use of operating system and hardware features as predictors to estimate the energy consumption at the component, virtual machine and application level  \cite{Dong_selfconstructivemodel, Kansal_powerofvm, Pathak_whereisenergy, Zhang_onlinepowerestimation, Wang_batterytrace, Shye_intowild}.

Shuai et al. \cite{HaoShuai:2013:EstMobileApp} and Ding et al. \cite{sourceline_energy} propose approaches to obtain source line energy information. The former requires the specific energy profile of the target system, and the workload is fine-tuned. The latter utilizes advanced measurement techniques to obtain the source line energy cost. 

Compared with approaches above, our latest work explores the idea of identifying energy operations and constructing a fine-grained model based on operations which is able to capture energy information at a level more fine-grained than source line. % \cite{Li_finemodel}
%\footnote{To preserve anonymity in the review process, the citation is hidden}
\paragraph{Energy-Saving Techniques}

%Li_smartcap,
A large amount of research effort on energy-saving for mobile devices has been focused on the main hardware components, such as the CPU, display and network interface. The CPU-related techniques involve dynamic voltage and frequency scaling \cite{anotherDVFS} and heterogeneous architecture \cite{Reflex_Lin, GreeDroid_Goulding}. Techniques targeting the display include dynamic back-light dimming \cite{dimming_backlight, dimming_backlight2} and tone-mapping based back light scaling \cite{tone_mapping,tone_mapping2}.  Network-related techniques try to exploit idle and deep sleep opportunities \cite{network1,network2}, shape the traffic patterns\cite{network_3,network4}, and so on. Such work attempts to reduce energy dissipation by optimizing the hardware usage; on the other hand, several pieces of work aim at designing new hardware and devices \cite{hardwaredisign1, hardwaredesign2}.

There is a significant research focus on software optimization for saving energy. The basic work seeks to understand how the different methods, algorithms and design patterns of software influence the energy consumption.
% The efforts in this part could be related to networking, memory and algorithms. 
For example, \cite{newRoutingTech1,newRoutingTech2,newRoutingTech3} propose new routing techniques and protocols that are aware of energy consumption, which are evaluated by comparing with traditional techniques.
For another example, \cite{choosesortalgorithm} investigates the affects of different sorting algorithms on the energy consumption with respect to the algorithm's input-size. 

Considering design patterns, Litke et al. \cite{desiangPattern1} conduct an experiment showing how big the difference of energy consumption is, before and after the application of design patterns, such as \textit{factory method pattern}, \textit{observer pattern}, and etc. The result reveals that except for one example the use of design patterns does not increase the energy use noticeably. Comparable work to \cite{desiangPattern1} is done by \cite{desiangPattern2};  they explores more design patterns and arrive at the conclusion that applying design patterns can both increase and decrease energy dissipation, so design-level artifacts cannot be used to estimate the impacts of design patterns on energy use.

Vetr\`{o} et al. \cite{code_smell} define the concept of "energy code smells" that are the code patterns (such as self assignment, repeated conditionals and useless control flow) that suggest energy inefficiency. However, the code patterns selected in \cite{code_smell} have very little influence (less than 1.0\%) on energy consumption.

% Our experimental result shows that our approach is able to save half of the entire energy consumption in certain scenario.   

Regarding code refactoring for energy-saving, Ding et al. \cite{energy_saving_programming} perform a small scale evaluation of several commonly suggested programming practices that may reduce energy. Its result shows that reading array length, accessing class field and method invocation all cost noticeable energy. However, this work only provides a small number of tips to developers on how to make the code more energy-efficient.

To the best of our knowledge,  the state of art before this paper did not connect the understanding of energy consumption with refactoring for such a high-level source code as Java. 
%no connection between understanding and optimization
%to the best of our knowledge 
Our research proposes an energy-aware programming approach, which is guided by the operation-based source-level energy model. The experimental evaluation demonstrates that the approach is an effective and practical approach to  energy-aware mobile application development.

%Moreover, our experiment is conducted on real application rather than 

%1) we construct the operation-based source-code-level energy model; 2) based on the model, we capture the energy characteristics of the code; 3) we improve the code by removing, reducing or replacing the expensive operations in the costly blocks.    
%correlating the operations to the energy cost by case analysis. Thus our model does not require  the profile of target system and more flexible to various pieces and types of code. The model is also capable to help produce the energy breakdown of the code to direct the developer's effort on energy efficiency.      

\section{Conclusion}

In this paper, we propose an energy-aware programming approach for mobile app development, guided by an operation-based source-level energy model. The approach consists of 1)  construction of an operation-based energy model by mining the data generated in a range of well-designed execution cases; 2) capturing energy characteristics of the code based on the model; 3) improving the code by removing, reducing or replacing the expensive operations in the costly blocks.   

We evaluate this approach on a physical Android development board with two ARM quad-core CPUs and on a real-world game engine. In this case study our approach has a significantly positive impact on energy-saving. For different scenarios, this approach can save energy between 6.4\% and 50.2\%. The findings also indicate that the performance of code is a potential by-product of this approach, which improves the user experience more.\\

%fine-grained energy model for mobile application source code on the basis of energy operations. We first introduce the energy operations that are identified directly from the source code. The energy operations are employed as the basic units that constitute the overall energy consumption of the source code. We then design a wide diversity of execution cases to generate data about the operation executions and the entire energy consumption. Regression analysis is applied to use the data to estimate the energy consumption of each operation. Finally, we show that the model is capable to capture comprehensive energy features that coarse-grained models or techniques could not shed light on.
%\section{Acknowledgements}

%This research is funded by the European Union Seventh Framework Programme (FP7/2007-2013) under grant agreement no 318337, ENTRA - Whole-Systems Energy Transparency.


\bibliographystyle{abbrv}
\bibliography{sigproc}  % sigproc.bib is the name of the Bibliography in this case

\end{document}